\documentstyle[12pt]{article}

\begin{document}
\setcounter{page}{1} \pagestyle{plain} \vspace{5cm}
\begin{center}
\large{\bf Entanglement witnesses for families of the bound entangled three-qubit states}\\
\small
\vspace{1cm} {\bf Y. Akbari-Kourbolagh}\\
\vspace{0.5cm} {Department of Physics,
Azarbaijan Shahid Madani University,\\
Tabriz 53741-161, Iran\\
yakbari@azaruniv.edu}
\end{center}
\vspace{1.5cm}
\begin{abstract}
We construct three-qubit entanglement witnesses with relatively simple structures. Despite their simplicity, these witnesses are capable of detecting a number of bound entangled states more effectively. To illustrate this, two families of bound entangled three-qubit states, introduced by Kay [Phys. Rev. A 83, 020303 (2011)] and Kye [J. Phys. A: Math. Theor. 48, 235303 (2015)], are considered and it is shown that these witnesses are able to detect the entanglement of all states in the families.

\end{abstract}
\newpage

\section{Introduction}
Detection of the entanglement in multi-qubit states is a fundamental issue in quantum information science. Multi-qubit entangled states are required in a wide range of quantum information processing tasks such as quantum teleportation \cite{aa,ab} and quantum cryptography \cite{1c}. The problem has been fully solved for two-qubit states \cite{1}-\cite{3}. For more than two qubits, however, the situation is more involved \cite{4}-\cite{10}. Even for the simplest case of three-qubit system, there exist six inequivalent types of entanglement \cite{4a}.
\par
There exist several different approaches to the detection of quantum entanglement. One of them which is of special interest, uses the notion of an entanglement witness \cite{20}. The entanglement witness is a Hermitian operator whose expectation values over all separable states are non-negative and there exist some entangled states for which the expectation value is negative. In the latter case, one says that the state is detected by the entanglement witness. The entanglement witness approach has its own advantages: it turns out that any entangled state is detected by some entanglement witnesses \cite{19} and the entanglement witness enables us to detect quantum states without having enough information about them.
\par
Here, following the method of \cite{7} and \cite{21}, we construct entanglement witnesses for three-qubit states. The witnesses have relatively simple structures, but they are nevertheless capable to detect a number of bound entangled states more effectively. To illustrate this, we consider the two families of bound entangled three-qubit states introduced by A. Kay \cite{16} and S. H. Kye \cite{18} and show that the witnesses are capable to detect the entanglement of all states in the families.
\par
It should be noted that our witnesses are members of a family of optimal linear entanglement witnesses which was introduced by Jafarizadeh et. al. \cite{22} for constructing nonlinear entanglement witnesses. Here, an alternative demonstration is offered for them and it is shown that they are more suitable for the detection of bound entanglement.
\par
The paper is structured as follows: In Section 2 we give a
brief description of the three-qubit states and the two families of bound entangled states introduced by A. Kay and S. H. Kye . In Section 3 we present the entanglement witnesses and apply them to the two families of bound entangled states. Finally, the paper is ended with our discussions and conclusion in section 4.
\section{Three-qubit states}
By definition, a three-qubit pure state $|\psi\rangle$ is separable or pure product if it can be written as a tensor product of three single qubit pure states, $|\psi\rangle=|\phi_{1}\rangle\otimes|\phi_{2}\rangle\otimes|\phi_{3}\rangle$. As for the three-qubit mixed state $\rho$, it is said to be separable if it can be expressed as a convex sum of separable pure states. Finally, a three-qubit mixed or pure state is said to be a PPT entangled or bound entangled state if it is entangled and its partial transposition on any qubit is positive.
\par
Here, we consider two families of the the three-qubit bound entangled states. The first family is the one introduced by A. Kay \cite{16}:
\begin{equation}\label{e2}
    \rho_{_{Kay}}(a):=\frac{1}{8+8a}\left(
      \begin{array}{cccccccc}
        4+a & 0 & 0 & 0 & 0 & 0 & 0 & 2 \\
        0 & a & 0 & 0 & 0 & 0 & 2 & 0 \\
        0 & 0 & a & 0 & 0 & -2 & 0 & 0 \\
        0 & 0 & 0 & a & 2 & 0 & 0 & 0 \\
        0 & 0 & 0 & 2 & a & 0 & 0 & 0 \\
        0 & 0 & -2 & 0 & 0 & a & 0 & 0 \\
        0 & 2 & 0 & 0 & 0 & 0 & a & 0 \\
        2 & 0 & 0 & 0 & 0 & 0 & 0 & 4+a \\
      \end{array}
    \right)
\end{equation}
Note that for any $a\geq 2$, the matrix $\rho_{_{Kay}}(a)$ is a valid quantum state and has a positive partial transposition for any bipartition. It was proved in \cite{16} that for $a\geq 2\sqrt{2}$ the state is separable. Furthermore, it was shown numerically that the state is entangled for $2\leq a\leq 2.828$. Finally, it was proved analytically that for $2\leq a< 2\sqrt{2}$ the state is entangled \cite{17}.
\par
The second family is the one introduced by S. H. Kye \cite{18}:
\begin{equation}\label{e2a}
    \rho_{_{Kye}}(b,c):=\frac{1}{6+b+c}\left(
      \begin{array}{cccccccc}
        1 & 0 & 0 & 0 & 0 & 0 & 0 & -1 \\
        0 & 1 & 0 & 0 & 0 & 0 & -1 & 0 \\
        0 & 0 & 1 & 0 & 0 & 1 & 0 & 0 \\
        0 & 0 & 0 & b & -1 & 0 & 0 & 0 \\
        0 & 0 & 0 & -1 & c & 0 & 0 & 0 \\
        0 & 0 & 1 & 0 & 0 & 1 & 0 & 0 \\
        0 & -1 & 0 & 0 & 0 & 0 & 1 & 0 \\
        -1 & 0 & 0 & 0 & 0 & 0 & 0 & 1 \\
      \end{array}
    \right)
\end{equation}
with $b$ and $c$ strictly positive real parameters such that $bc\geq 1$. It is clear that the state is of PPT  whenever $bc=1$.
  \section{Entanglement witnesses}
To construct entanglement witnesses for the bound entangled states $\rho_{_{Kay}}(a)$ and $\rho_{_{Kye}}(b,c)$, let us consider the observables (Hermitian operators)
\begin{equation}\label{e3}
    A_{1}=\sigma_{z}\otimes\sigma_{z}\otimes I,\; A_{2}=\sigma_{x}\otimes(\sigma_{x}\otimes\sigma_{x}-\sigma_{y}\otimes\sigma_{y})
    ,\;A_{3}=\sigma_{y}\otimes(\sigma_{x}\otimes\sigma_{y}-\sigma_{y}\otimes\sigma_{x}),
\end{equation}
where $I$ is the single qubit identity operator and $\sigma_{i}, i=x,y,z$ are the usual single qubit Pauli operators.
Now, if any three-qubit state $\rho$ is characterized by the expectation values of the operators $A_{i}$ over it $(\langle A_{1}\rangle_{\rho}, \langle A_{2}\rangle_{\rho}, \langle A_{3}\rangle_{\rho} )$, where $\langle A_{i}\rangle_{\rho}:=\mathrm{Tr}(\rho A_{i})$, then we have the following observation.
\par
{\bf{ Observation}:} For any separable three-qubit state $\sigma$, the following inequality is fulfilled
\begin{equation}\label{e4}
    \langle A_{1}\rangle_{\sigma}+\frac{1}{\sqrt{2}}(\langle A_{2}\rangle_{\sigma}+\langle A_{3}\rangle_{\sigma})\leq1.
\end{equation}
The inequality may be violated by entangled states.
\par
Obviously the observation remains valid when the observables $A_{i}$ are replaced with the ones that come from all the possible permutations of $\sigma_{x}, \sigma_{y}$ and $\sigma_{z}$ and also when they are replaced with $-A_{i}$.
\par
\textbf{Proof:} As the expectation value of an operator is linear in the state, it suffices to prove that the
inequality (\ref{e4}) is satisfied by pure product states. For the pure product state $|\psi\rangle=|\phi_{1}\rangle\otimes|\phi_{2}\rangle\otimes|\phi_{3}\rangle$, we have
\begin{equation}\label{e5}
\begin{array}{c}
\hspace{-3.7cm}\langle A_{1}\rangle_{|\psi\rangle}+\frac{1}{\sqrt{2}}(\langle A_{2}\rangle_{|\psi\rangle}+\langle  A_{3}\rangle_{|\psi\rangle})=
 \langle\phi_{1}|\sigma_{z}|\phi_{1}\rangle\langle\phi_{2}|\sigma_{z}|\phi_{2}\rangle\\
  +\frac{1}{\sqrt{2}}\langle\phi_{1}|\sigma_{x}|\phi_{1}\rangle(\left\langle\phi_{2}|\sigma_{x}|\phi_{2}\rangle
  \langle\phi_{3}|\sigma_{x}|\phi_{3}\rangle-\langle\phi_{2}|\sigma_{y}|\phi_{2}\rangle
  \langle\phi_{3}|\sigma_{y}|\phi_{3}\rangle\right)\\
\hspace{.1cm}+\frac{1}{\sqrt{2}}\langle\phi_{1}|\sigma_{y}|\phi_{1}\rangle(\langle\phi_{2}|\sigma_{x}|\phi_{2}\rangle
  \langle\phi_{3}|\sigma_{y}|\phi_{3}\rangle-\langle\phi_{2}|\sigma_{y}|\phi_{2}\rangle
  \langle\phi_{3}|\sigma_{x}|\phi_{3}\rangle).
\end{array}
\end{equation}
Now let us introduce the notation
\begin{equation}\label{e6}
    x_{i}:=\langle\phi_{i}|\sigma_{x}|\phi_{i}\rangle \quad,\quad y_{i}:=\langle\phi_{i}|\sigma_{y}|\phi_{i}\rangle \quad,\quad z_{i}:=\langle\phi_{i}|\sigma_{z}|\phi_{i}\rangle.
\end{equation}
Then Eq. (\ref{e5}) is rewritten as
\begin{equation}\label{e7}
    \langle A_{1}\rangle_{|\psi\rangle}+\frac{1}{\sqrt{2}}(\langle A_{2}\rangle_{|\psi\rangle}+\langle  A_{3}\rangle_{|\psi\rangle})=z_{1}z_{2}+\frac{1}{\sqrt{2}}[x_{1}(x_{2}x_{3}-y_{2}y_{3})
    +y_{1}(x_{2}y_{3}-y_{2}x_{3})].
\end{equation}
By the well-known fact that  $x_{i}^{2}+y_{i}^{2}+z_{i}^{2}=1$, it is convenient to use the parametrization
\begin{equation}\label{}
     x_{i}:=\sin\theta_{i}\cos\varphi_{i} \quad,\quad y_{i}:=\sin\theta_{i}\sin\varphi_{i} \quad,\quad z_{i}:=\cos\theta_{i},
\end{equation}
where $\theta_{i}\in[0, \pi]$ and $\varphi_{i}\in[0, 2\pi]$ are real parameters. With respect to this parametrization, Eq. (\ref{e7}) takes the form
\begin{equation}\label{e8}
\begin{array}{c}
\hspace{-5.5cm}\langle A_{1}\rangle_{|\psi\rangle}+\frac{1}{\sqrt{2}}(\langle A_{2}\rangle_{|\psi\rangle}+\langle  A_{3}\rangle_{|\psi\rangle})=\cos\theta_{1}\cos\theta_{2} \\
  \hspace{.4cm}+\frac{1}{\sqrt{2}}\sin\theta_{1}\sin\theta_{2}\sin\theta_{3} [\cos\varphi_{1}\cos(\varphi_{2}+\varphi_{3})-\sin\varphi_{1}\sin(\varphi_{2}-\varphi_{3})].
\end{array}
\end{equation}
Obviously, the right hand side of Eq. (\ref{e8}) can not exceed 1 and takes the maximum value 1 for $\varphi_{3}=-\varphi_{2}=\frac{\pi}{4}$, $\varphi_{1}=\frac{\pi}{4}$, $\theta_{1}=\theta_{2}$, and $\theta_{3}=\frac{\pi}{2}$.
\par
The observation provides a necessary criterion for the separability of three-qubit states whose violation is a sufficient criterion for the entanglement. Furthermore, we infer from the inequality (\ref{e4}) that the observable
\begin{equation}\label{e9}
     W_{1}:=I-A_{1}-\frac{1}{\sqrt{2}}(A_{2}+A_{3})
\end{equation}
with $I$ the three-qubit identity operator, has a non-negative expectation value over any three-qubit separable state and therefore it is an entanglement witness.
\par
To see the detection power of $W_{1}$, let us consider the state $\rho_{_{Kay}}(a)$ of (\ref{e2}).
For this state, we have
$$
{\mathrm{Tr}}(W_{1}\rho_{_{Kay}}(a))=1-\frac{2\sqrt{2}+1}{a+1}.
$$
When this expectation value is negative it is said that $\rho_{_{Kay}}(a)$ is detected by $W_{1}$. Obviously, the expectation value is negative for $2\leq a< 2\sqrt{2}$ and therefore for these values of $a$ the state $\rho_{_{Kay}}(a)$ is entangled. This result is in excellent agreement with the one of \cite{17}.
\par
As another example, let us consider the state $\rho_{_{Kye}}(b,c)$ of (\ref{e2a}). For this state, we use the following witness
\begin{equation}\label{e10}
     W_{2}:=I+A_{1}+\frac{1}{\sqrt{2}}(A_{2}+A_{3})
\end{equation}
which is inferred from the observation when the $A_{i}$'s are replaced with $-A_{i}$'s.
For this state, we have
$$
{\mathrm{Tr}}(W_{2}\rho_{_{Kye}}(b,c))=-\frac{8(\sqrt{2}-1)}{6+b+c}.
$$
Obviously this expectation value is negative for any allowed values of $b$ and $c$. Therefore, the state $\rho_{_{Kye}}(b,c)$ is detected by $W_{2}$ and therefore it is entangled.

\section{Discussions and conclusion}
We present entanglement witnesses for the entanglement of three-qubit states and find that, despite the simplicity, they are relatively strong in detecting a number of bound entangled states. We show that they detect all the states in the families of bound entangled states introduced by A. Kay and S. H. Kye. It would be interesting to apply the method to other three-qubit bound entangled states and extend it to more than three qubits systems.

\par
{\bf Acknowledgements:} We acknowledge very valuable discussions with M. A. Jafarizadeh
and also thanks M. A. Fasihi for the interesting and useful
discussions.

\end{document}